\documentclass[12pt]{iopart}

\usepackage[colorlinks,citecolor=blue,linkcolor=red,anchorcolor=blue,filecolor=blue,urlcolor=blue]{hyperref}
\usepackage{color}
\usepackage{graphicx}

\begin{document}

\title[Isotopic shift in charge radii across $N$ = 82 and 126 revisited]{The kinks in charge radii across $N$ = 82 and 126 revisited}

\author{M. Bhuyan$^{1,2}$,B. Maheshwari$^{3}$, H. A. Kassim$^1$, N. Yusof$^1$, S. K. Patra$^{4,5}$, B. V. Carlson$^6$, P. D. Stevenson$^7$ }

\bigskip
\address{$^1$Center for Theoretical and Computational Physics, Department of Physics, Faculty of Science, University of Malaya, Kuala Lumpur 50603, Malaysia \\
$^2$Institute of Research and Development, Duy Tan University, Da Nang 550000, Vietnam \\
$^3$Amity Institute of Nuclear Science and Technology, AUUP, Noida 201313, India \\
$^4$Institute of Physics, Sachivalaya Marg, Sainik School, Bhubaneswar 751005, India \\
$^5$Homi Bhabha National Institute, Training School Complex, Anushakti Nagar, Mumbai 400085, India \\
$^6$Instituto Tecnol\'ogico de Aeron\'autica, S\~ao Jos\'e dos Campos 12.228-900, S\~ao Paulo,Brazil \\
$^7$Department of Physics, University of Surrey, Guildford GU2 7XH, United Kingdom}
\ead{bunuphy@um.edu.my}
\ead{bhoomika.physics@gmail.com} 
\ead{p.stevenson@surrey.ac.uk}
\vspace{10pt}
\begin{indented}
\item[]February 2021
\end{indented}

\begin{abstract}
\noindent 
We revisit the studies of the isotopic shift in the charge radii of {\it even-even} isotopes of Sn and Pb nuclei at $N$ = 82, and 126, respectively, within the relativistic mean-field and Relativistic-Hartree-Bogoliubov approach. The shell model is also used to estimate isotopic shift in these nuclei, for the first time, to the best of our knowledge. The ground state single-particle energies ($spe$) are calculated for non-linear NL3 \& NL3$^*$ and density-dependent DD-ME2 parameter sets compared with the experimental data, wherever available. We establish a correlation between the filling of single-particle levels and the isotopic shift in occupation probabilities. The obtained $spe$ from the relativistic mean-field and Relativistic-Hartree-Bogoliubov approaches are in line with those used in the shell model and experimental data for both the Sn and Pb isotopic chains. The shell model calculated isotopic shift agrees with relativistic mean-field and Relativistic-Hartree-Bogoliubov approaches that explain the experimental data quite well. In case of Pb nuclei beyond N=126, the shell model results stay far from the data.   
\end{abstract}
\maketitle

\section{Introduction}
\label{intro}
In nuclear systems, one of the most important degrees of freedom is the isospin asymmetry, which presents an active research field. The isospin asymmetry is well connected with the nuclear symmetry energy of infinite nuclear matter and the neutron skin thickness of finite nuclei, which play an important role in understanding many issues in nuclear physics \cite{dani02,stei05,bara05,latt07,li08,dong11,bhu15} as well as in nuclear astrophysics \cite{latt04,horo01,agra12,wan17,bhu18}. Following the available information on neutron radii, isotopic shift, and neutron skin thickness, one can find notable uncertainties in the empirical constraints as shown by Sammarruca \cite{samm16} and references therein. In the cases of $^{48}$Ca and $^{208}$Pb, a variety of measurements have been undertaken \cite{tsan12,garc92,frie12,gibb92} and a few future experiments \cite{abra12}, are planned to probe these uncertainties in the charge density. The possibility of obtaining reliable values for neutron or proton skins is hindered by limitations analogous to the charge density. The measurements of both proton and neutron radii must be done with great accuracy to extract either skin thickness and/or isotopic shift, which has proved to be quite important for constraining the predictability of nuclear models at high isospin asymmetry. 

Nuclear charge radii have been studied theoretically within various microscopic-macroscopic ({\it mic-mac}) models \cite{thib81,muel83}, which cover the basic idea of their evolution over the nuclear chart. In some specific cases, the quantum shell effects dominate the charge distribution and lead to a deviation from smooth systematic trends \cite{camp16}. Perhaps the most well-known example of this abrupt change is in the isotope shift (kink) of even-even Pb isotopes at $N = 126$. Such kinks are not limited to the $Z = 82$ region, but also known in nuclei with 50$\leq Z \leq$ 82 and 82$ \leq N \leq$ 126 \cite{godd13,krei14,ha18,fu18,rein17,gorg19}. This gives rise to the question of whether the knowledge of neutron (proton) distributions alone can provide information for the neutron (proton) skin-thickness and isotopic shift. A large number of studies have tried to explain the kink simultaneously at $N = 82$ and $126$ for Sn- and Pb- isotopes, respectively \cite{godd13,rein17,coco11,taji93,faya00,tolo10,sape11,verm92,garc16}. Most of the non-relativistic nuclear energy density functionals (EDFs) are not able to produce the kink at $N = 82$ for Sn isotopes. On the other hand, beyond mean-field treatments such as an extended pairing-field on density-dependent spin-orbit interactions within various semi-realistic models are found to be competent to some extent \cite{faya00,tolo10,naka15}. 

In the structural properties of $\beta$-stable $^{208}$Pb and neutron-rich $^{132}$Sn, one can find a few similarities in terms of an average large shell-gap, being double-magic, with zero pairing-gap and so on \cite{borc89,jone10,agbe14,bhu12,afan16}. The main issue is a consistent explanation of the neutron skin, and isotopic shift for both the Sn and Pb isotopic chains within a particular model \cite{faya00,verm92,borc89,bhu15,bhu18}. Recently, additional structural information has been obtained for the doubly-magic isotope $^{132}$Sn from the High-Intensity and Energy upgrade of ISOLDE (HIE-ISOLDE) \cite{rosi18}, laser spectroscopic and muonic measurements of nuclear charge radii \cite{bair83,eber87,blan05} and from RIKEN \cite{yasu18}. On the other hand, there are various theoretical studies that have been performed to simultaneously predict the charge radii of highly isospin asymmetric isotopes, especially $^{36}$Ca \cite{mill19}, $^{52}$Ca \cite{garc16}, $^{52,53}$Fe \cite{mina16}, and $^{100-130}$Cd \cite{hamm16}. The high-precision laser spectroscopic measurements of the charge radii of $^{108-134}$Sn \cite{gorg19}, and the theoretical explanation of the isotopic shift at $N = 126$ of Pb- and Po-isotopes \cite{godd13} motivate us to perform a systematic study of the isotopic shift of Sn and Pb isotopes at $N = 82$ and 126, respectively. In general, the kinks in charge radii over the isotopic chain, i.e., isotope shifts, are related to the evolution of the single-particle energy levels at/near the magic neutron numbers. The idea of magicity was originally introduced by the shell model \cite{mayer55}. Hence, it will be interesting to interlink the single-particle energies and the isotopic shift in the charge radii using the shell model in parallel with the relativistic mean-field approach. In the present analysis, we use relativistic mean-field  (RMF) formalism with the very popular NL3 and the recently developed NL3$^*$, and Relativistic-Hartree-Bogoliubov (RHB) with DD-ME2 parameter sets to estimate the isotopic shift of the Sn- and Pb- isotopic chains. We have also performed the shell model calculations, for the first time (to the best of our knowledge), to obtain the isotopic shift for both the Sn and Pb isotopic chains at and around $N=82$ and $126$, respectively.

The paper is organized as follows: Sec. \ref{theory} gives a brief description of the RMF approach for single-particle distributions. The effects of pairing for open-shell nuclei within the RMF are also discussed in this section. The details of RMF and shell model calculations along with results will be presented in Sec. \ref{result}. A brief summary of the results obtained, together with concluding remarks, is given in Sec. \ref{summary}.

\section{Theoretical Formalism}
\label{theory}
In the last few decades, the relativistic mean-field and Relativistic-Hartree-Bogoliubov models have been applied successfully to study the structural properties of nuclei throughout the nuclear chart, including drip-line nuclei and the exotic island of superheavy nuclei \cite{sero86,boguta77,horo81,boguta83,patra91,ring86,niks02,lala05,bhu18,bhu20a}. The documentation of the relativistic Lagrangian density for many-body nucleon-meson system and the standard expressions for bulk properties such as binding energy, quadrupole deformation, root-mean-square radius, density distributions, and single-particle energies can be found in Refs. \cite{boguta83,ring86,boguta77,horo81,patra91,ring86,bhu09,bhu11,bhu15,bhu18,ring90,niks02,lala05,bhu20}. We use the recently developed non-linear NL3$^*$ \cite{lala09} and density-dependent DD-ME2 \cite{lala05} for the present analysis. The shell model calculation for the isotopic shift in the charge radii is also included in the present studies. The obtained results are compared with the well-established NL3 parameter set \cite{lala97}. It is worth mentioning that the RMF is considered to be one of the most successful models to reproduce the ground state properties not only for $\beta-$stable nuclei but also to predict reasonably well the properties of drip-line and superheavy nuclei \cite{patra91,ring86,bhu09,bhu11,bhu15,bhu18,rutz95,rutz98,book01,bhu20}.

To describe the nuclear bulk properties of open-shell nuclei, one must consider the pairing correlations in their ground states. There are various methods, such as the BCS approach, the Bogoliubov transformation, and particle number conserving methods, that have been developed to treat pairing effects in the study of nuclear properties, including fission barriers \cite{zhan11,hao12,bhu15,bhu18}. In principle, the Bogoliubov transformation is the most widely used method to take pairing correlations into account for the drip-line region \cite{niks02,lala05,bhu18,bhu20a}. In the case of nuclei not too far from the $\beta$-stability line, one can use the constant gap BCS pairing approach to obtain a reasonably good approximation of pairing \cite{doba84}. We have employed the constant gap BCS approach with the NL3 and NL3$^*$ and a Bogoliubov transformation with DD-ME2 parameter sets in the present studies. We are dealing with the Sn and Pb isotopes near the $\beta$-stable region, which are almost spherical. Hence, the constant gap BCS-pairing and Bogoliubov transformation approaches are valid in present calculations \cite{madland81}. Many other authors and we have already applied these kinds of prescription for pairing correlations both in the RMF and Skyrme-Hartree-Fock (SHF) \cite{patra01,bhu11,bhu11a,bhu15,bhu18,bhu19,niks02,lala97,lala09,lala05}.

\begin{figure}
\begin{center}
\includegraphics[width=0.67 \columnwidth]{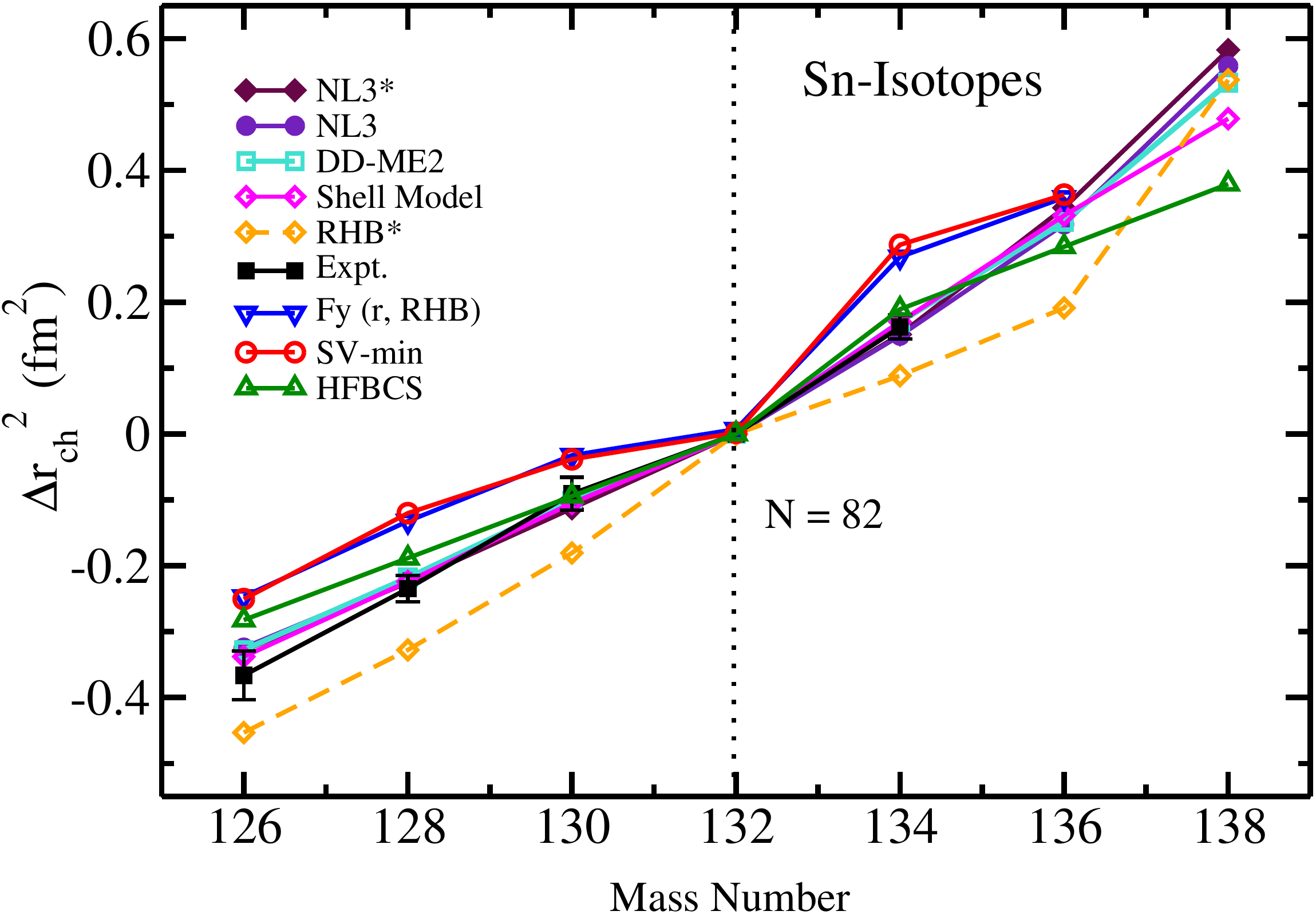}
\includegraphics[width=0.67 \columnwidth]{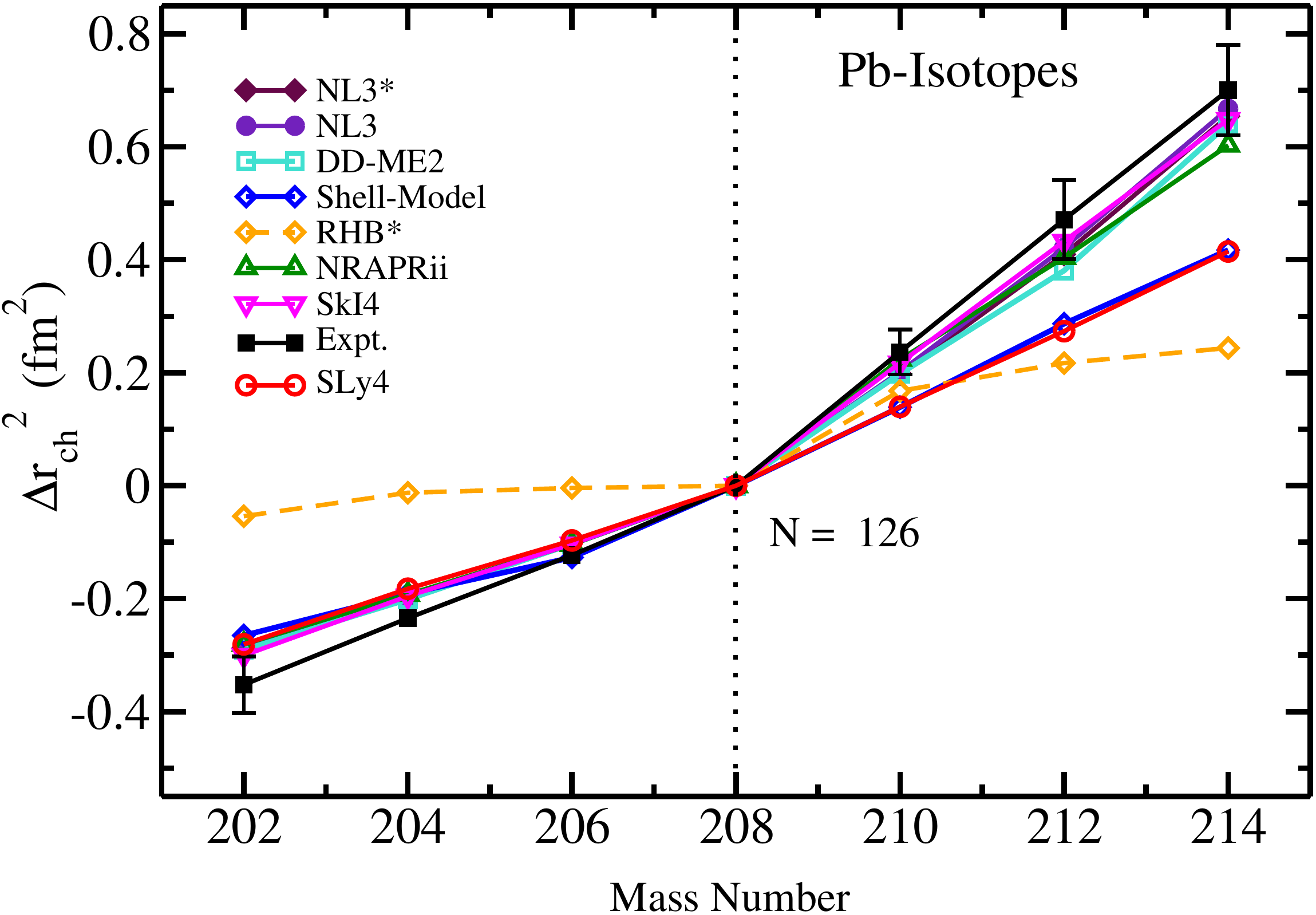}
\caption{\label{fig1} (Color online) The isotopic shift over the isotopic chains of Sn- (upper panel) and Pb-nuclei (lower panel) within the relativistic mean-field formalism for NL3 and NL3$^*$, Relativistic-Hartree-Bogoliubov for DD-ME2 parameter sets, and shell model calculations are shown. The RHB$^*$ stands for the calculations using occupancies from Relativistic-Hartree-Bogoliubov for DD-ME2 parameter set. Other theoretical predictions \cite{gori01,rein17,godd13} (references therein) along with the experimental data \cite{fric04,angeli13,gorg19} (references therein) are given for comparison. See the text for more details.}
\end{center}
\end{figure}
\begin{figure}[ht]
\begin{center}
\includegraphics[width=0.6 \columnwidth]{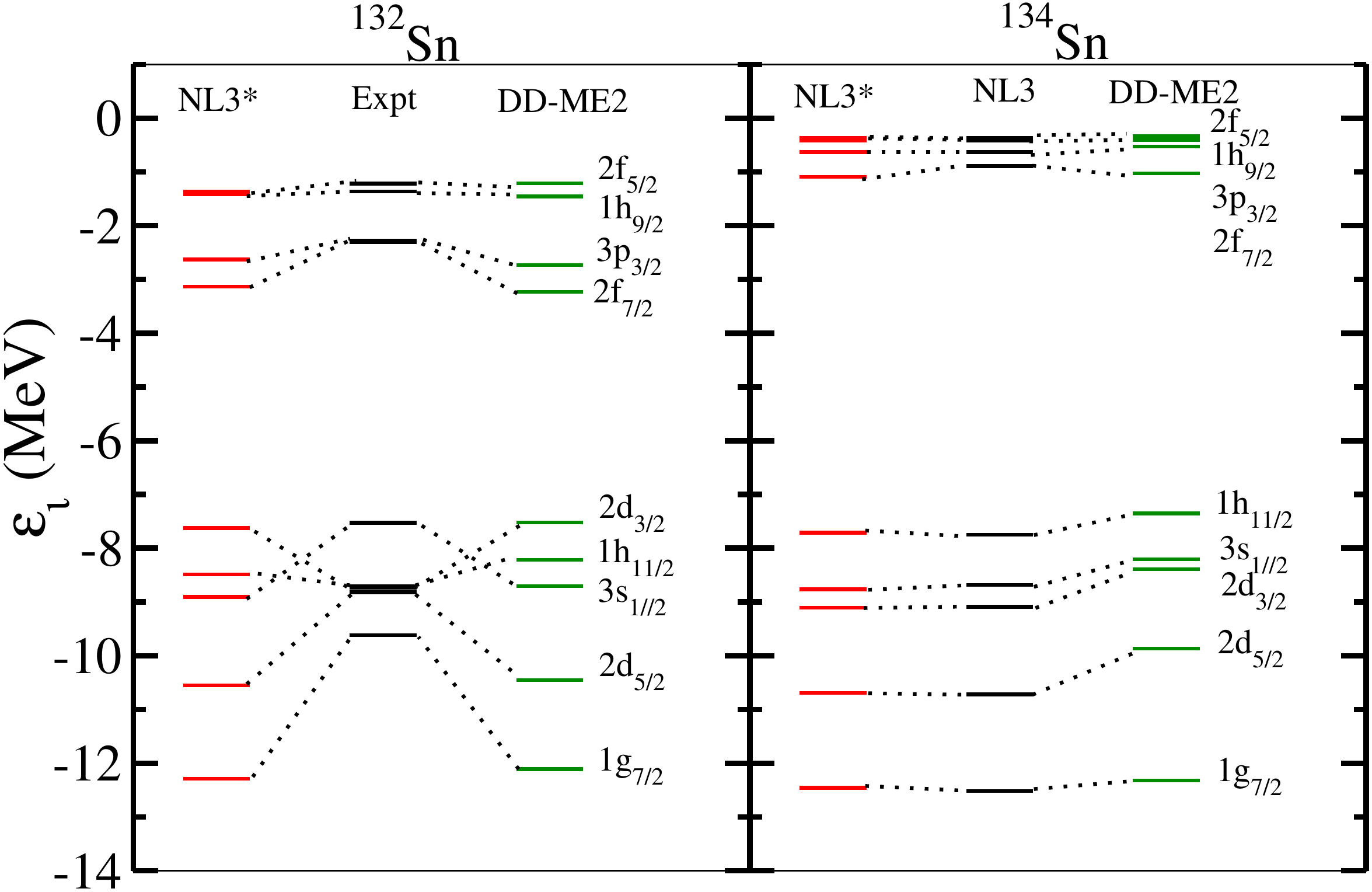}
\includegraphics[width=0.6 \columnwidth]{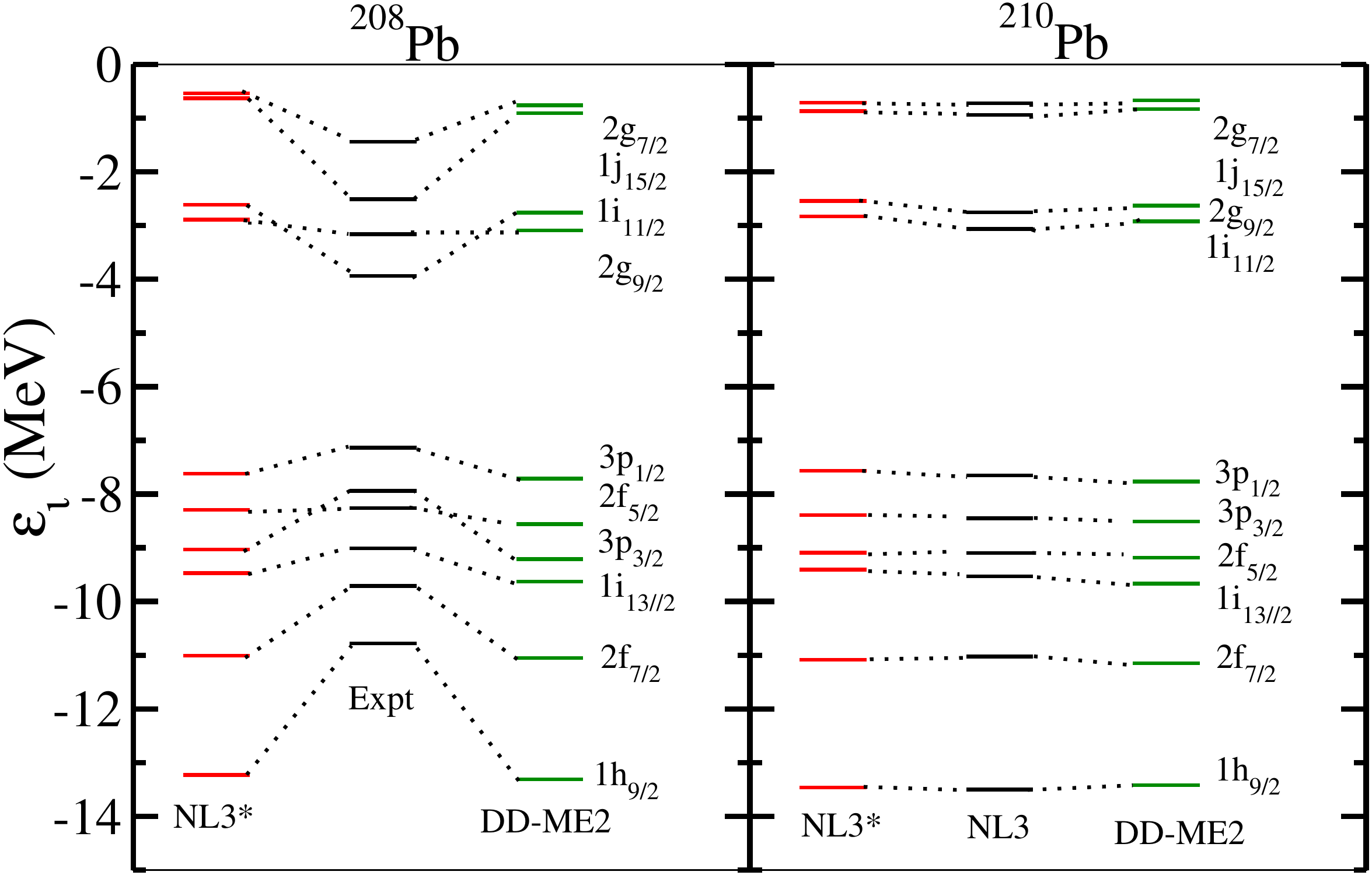}
\includegraphics[width=0.6 \columnwidth]{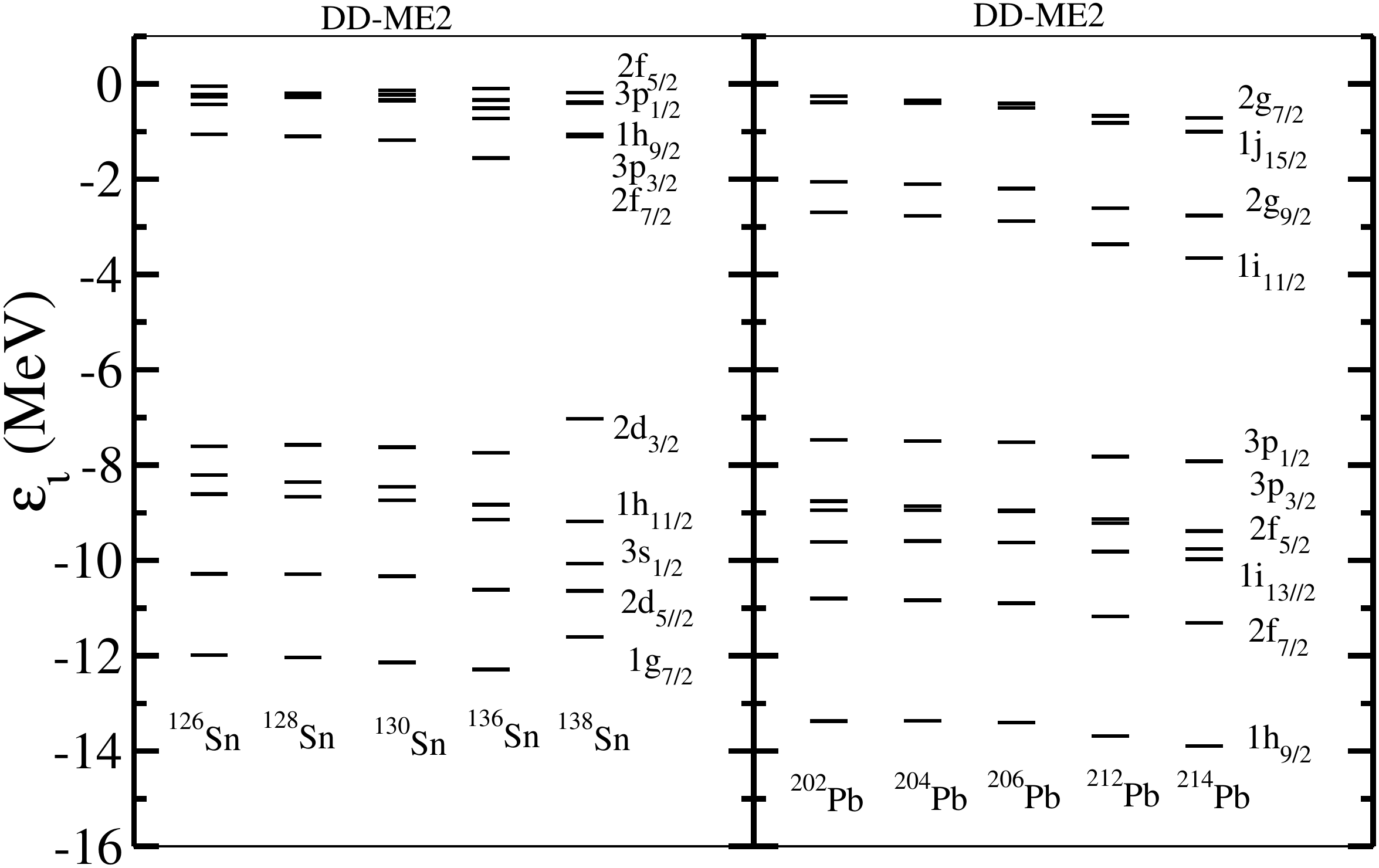}
\caption{\label{fig2} (Color online) The single-particle energies near the Fermi level for $^{132}$Sn (left upper panel) and $^{208}$Pb (left middle panel) within the relativistic mean-field formalism for NL3$^*$ and Relativistic-Hartree-Bogoliubov for DD-ME2 parameter sets are compared with experimental data \cite{brow98}. The right upper and middle panels are for the respective single-particle energies of $^{134}$Sn, and $^{210}$Pb, for both NL3, NL3$^*$, and DD-ME2 parameter sets. The single-particle energy for the rest of the neutron numbers for Sn- and Pb- isotopes are shown in the left and right lower panel for the DD-ME2 parameter set.}
\end{center}
\end{figure}

\section{Calculations and Results}
\label{result}
A large number of interaction parameter sets exist within the relativistic mean-field approach for finding a convergent and self-consistent solution for the ground and intrinsic excited states of finite nuclei. In many of our previous works and of others \cite{patra91,ring90,lala97,lala09,bhu09,bhu11,bhu11a,bhu12,bhu15,bhu18,bhu19}, the ground-state properties such as the binding energies (BE), quadrupole deformation parameters ($\beta_2$), root-mean-square charge radii ($r_{ch}$), single-particle levels ($\epsilon_{n,p}$), nuclear distributions ($\rho_{n,p}$) and other properties, are being evaluated by using the various parameter sets. From these works, one can conclude that most of the recent parameter sets describe ground-state properties reasonably well, not only for stable nuclei but also for exotic drip-line nuclei, including superheavy nuclei \cite{bhu09,bhu11,bhu11a,bhu12,bhu15,bhu18,bhu19}. This implies that the predictions from a recent acceptable parameter set will remain nearly force-independent. In this context, we have used the recently developed NL3$^*$ \cite{lala09}, and density-dependent DD-ME2 \cite{lala05} parameter sets for the present investigation and also obtained results using the popular NL3 \cite{lala97} parameter for standard comparison. The numerical calculations are carried out by taking the maximum oscillator quanta $N_F$ = 16 for fermions. To test the convergence of the solutions, we vary the $N_B$ (number of oscillator quanta for bosons) from 12 to 18. The variation of these solutions is $\leq$ 0.05$\%$ and 0.02$\%$ in binding energy and root-mean-square matter radius, respectively. Hence, the used model space $N_F$ = $N_B$ = 16 is good enough for the considered nuclei. The number of mesh points for the Gauss-Hermite and Gauss-Laguerre integration is 20 and 24, respectively. For a given nucleus, the maximum binding energy corresponds to the ground state, and other solutions are considered intrinsic excited states. The ground-state rms charge radii and the single-particle energies for Sn- and Pb-isotopes are obtained using the RMF with NL3 and NL3$^*$ and Relativistic-Hartree-Bogoliubov with DD-ME2 parameter sets.  \\

\noindent
{\bf Isotopic shift at $\mathbf{N = 82}$ and 126:}
In this subsection, we present the relative change in the root-mean-square charge radius in terms of the isotopic shift for $^{126-140}$Sn and $^{202-216}$Pb nuclei. The isotopic shift at a reference neutron number can be defined as , 
\begin{eqnarray}
\langle \Delta r_{ch}^2 \rangle^{A} = \langle  r_{ch}^2 \rangle^{A} 
-\langle  r_{ch}^2 \rangle^{i}, 
\label{shift}
\end{eqnarray}
where $i$ stands for the mass corresponding to the neutron number 82 and 126 for Sn- and Pb- nuclei, respectively. From the calculated r$_{ch}$ values, we obtain the isotopic shift $\langle \Delta r_{ch}^2 \rangle^{A}$ using Eq. (\ref{shift}) for NL3, NL3$^*$, DD-ME2 parameter sets, and shell model for both the Sn- and Pb- isotopic chains. The results for the $\langle \Delta r_{ch}^2 \rangle^{A}$ along with the experimental data \cite{fric04,angeli13,gorg19,raman01} and other theoretical predictions \cite{gori01,rein17,godd13} (references therein) are displayed in Fig. \ref{fig1}. The upper and lower panel of Fig. \ref{fig1} are given for Sn- and Pb-isotopes, respectively. We notice a clear shift at $N$ = 82 and 126, respectively, for Sn- and Pb- isotopic chains. This is in agreement with the experimental data. The strength of the isotopic shift at $N$ = 82 for Sn- is weaker as compared to $N$ = 126 for Pb-nuclei. This can be correlated with the soft properties of Sn-isotopes \cite{biswal15,howa20} and references therein. A quantitative analysis (of our calculations from NL3, NL3$^*$, DD-ME2 parameter sets) shows that the isotopic shift for Sn-isotopes fits a slope of 0.035 $\pm$ 0.015, and 0.105 $\pm$ 0.018 for mass A $\leq$ 132 and for A $\geq$ 132, respectively. Similarly, the slopes in the case of the Pb-nuclei fit 0.047 $\pm$ 0.015, and 0.116 $\pm$ 0.018 for mass A $\leq$ 208, and for A $\geq$ 208, respectively. Recent studies \cite{godd13,krei14,ha18,fu18,rein17,gorg19} have shown that the isotopic shift can be directly connected with the single-particle energy levels and the filling of these orbits. In other words, the occupation probabilities for the neutrons are crucial to determine the isotopic shift over an isotopic chain. This will be discussed subsequently in the next paragraph.

\begin{figure}[!ht]
\centering \includegraphics[trim=75 25 100 45, clip,width=0.7\columnwidth]{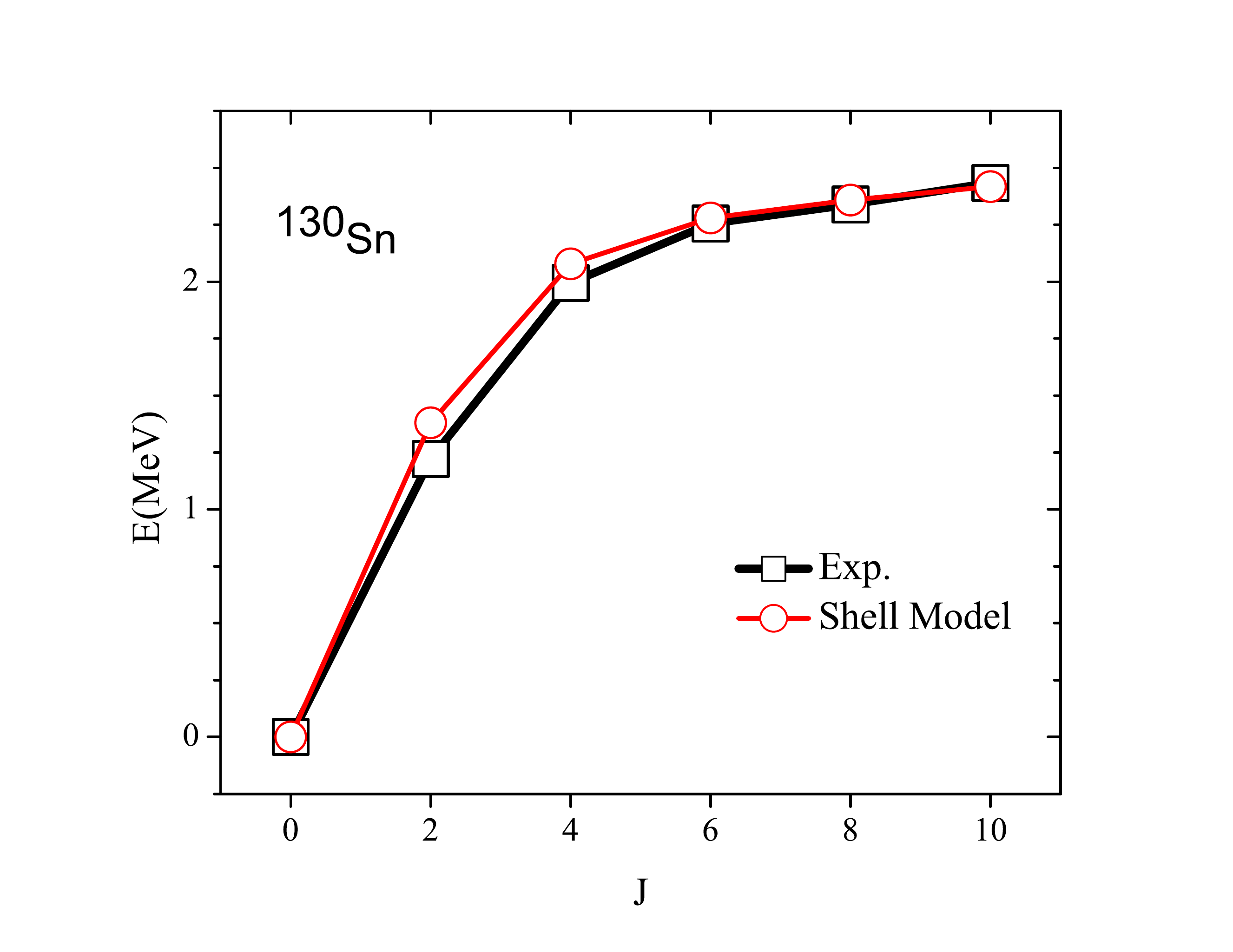}
\centering \includegraphics[trim=75 25 100 45, clip,width=0.7\columnwidth]{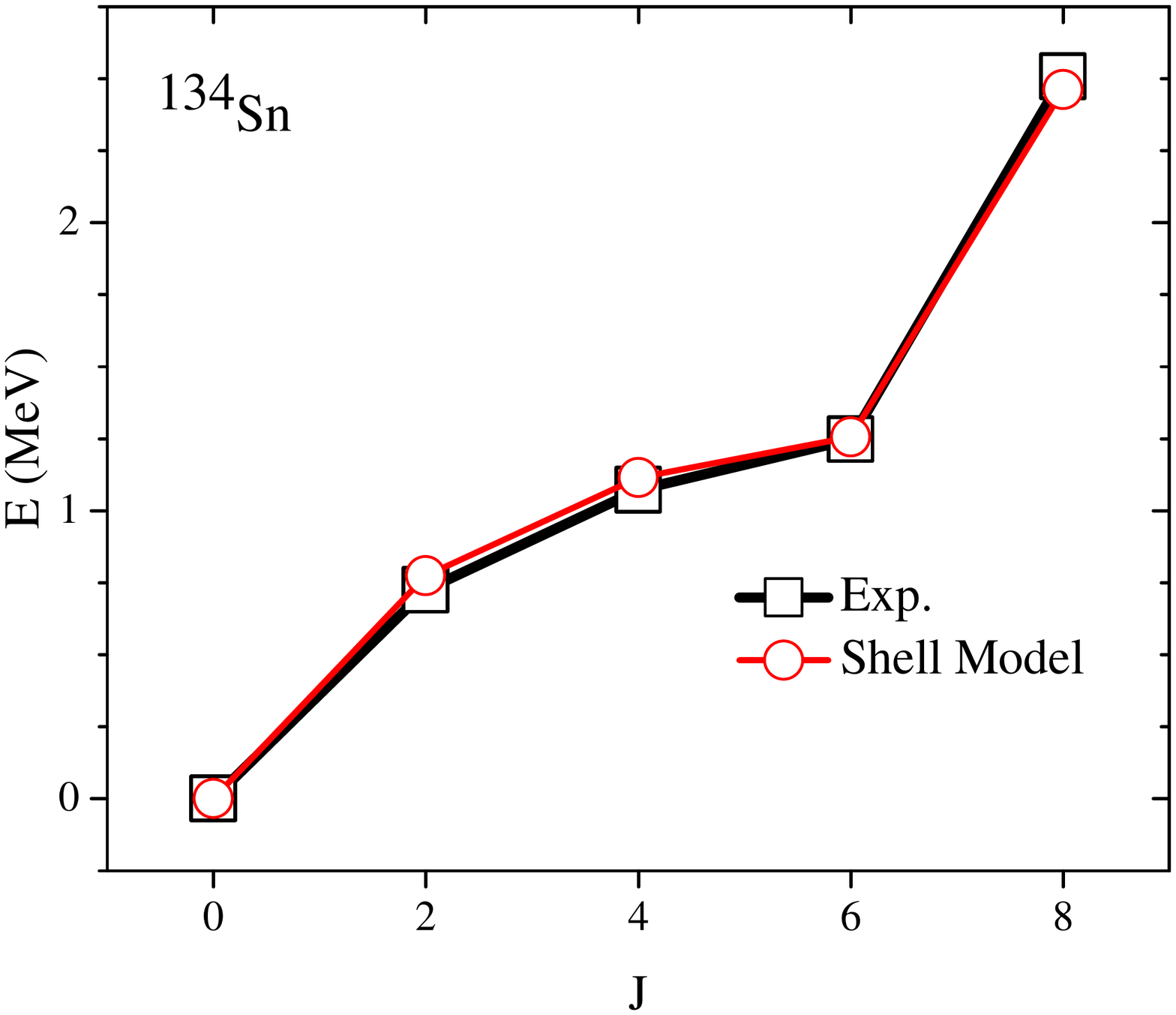}
\caption{\label{fig3}(Color online) Experimental \cite{nndc} and Shell model calculated energies for yrast $0^+$ to ${10}^+$ states in $^{130}$Sn and for yrast $0^+$ to $8^+$ states in $^{134}$Sn.}
\end{figure}
\begin{figure}[!ht]
\begin{center}
\includegraphics[trim=75 20 100 65, clip,width=0.7\columnwidth]{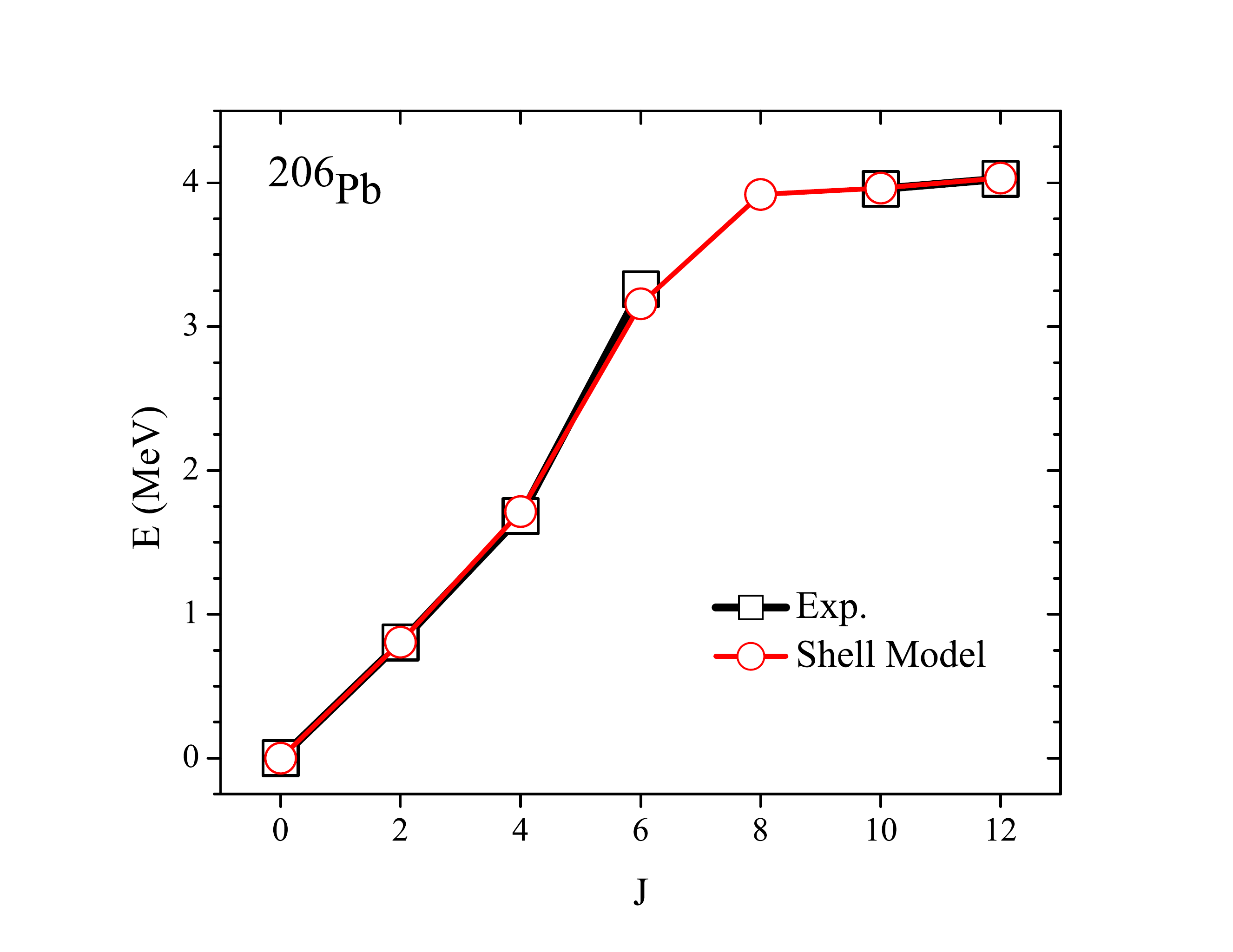}
\includegraphics[trim=75 25 100 65, clip,width=0.7\columnwidth]{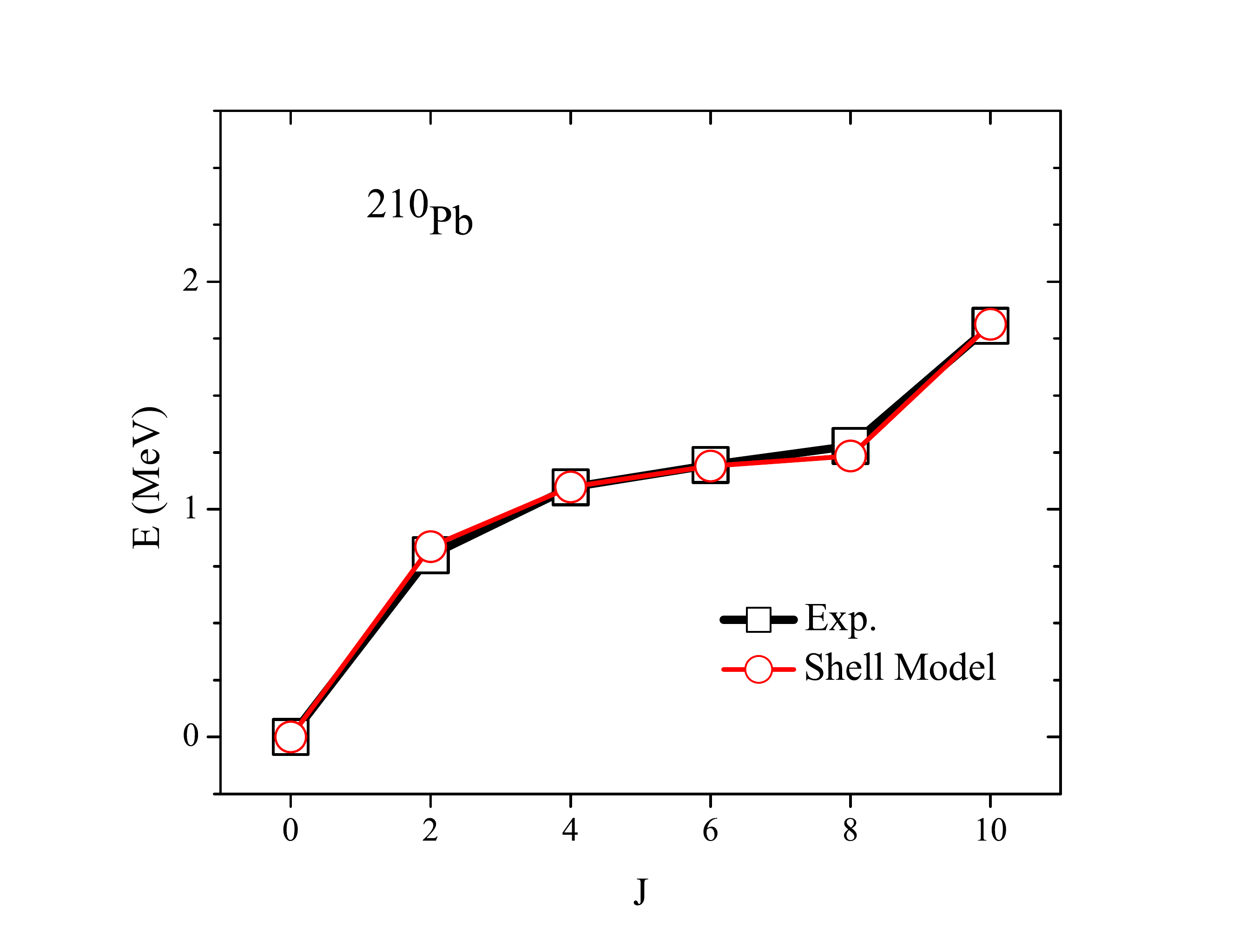}
\caption{\label{fig4}(Color online) Same as for Fig. 3, but for the yrast $0^+$ to ${12}^+$ states in $^{206}$Pb and for the yrast $0^+$ to ${10}^+$ states in  $^{210}$Pb.}
\end{center}
\end{figure}

\noindent
{\bf Single-particle energies:}
The single-particle energies (SPEs) are one of the important facets for understanding the nuclear structure. They play a role in explaining the magicity (shell/sub-shell closures) and directly influence bulk properties such as the nuclear radii and the matter density-distributions of the nucleons. This paper mainly concentrates on the neutron single-particle energies since we are dealing with the isotopic chains of Sn- (Z = 50) and Pb- (Z = 82) nuclei. We present RMF (NL3$^*$) (red lines), RHB (DD-ME2) (blue line) neutron single-particle energy levels for $^{132}$Sn and $^{208}$Pb in the left upper and middle panels of Fig. \ref{fig2}, respectively. Similar trends and almost the same order of energy values can be found for the NL3 parameter set, which is not shown to maintain the clarity in Fig. \ref{fig2}. The levels near the Fermi energy are given in Fig. \ref{fig2} along with the experimental data (black line) \cite{brow98} and references therein. Regarding the magnitude of SPEs, a good match between the experimental data and the theoretical calculations can be observed for the hole (particle) valence levels of the selected closed-shell nuclei. For example, the energies for the three shallowest hole (particle) states, namely, 3$s_{1/2}$, 1$h_{11/2}$, and 2$d_{3/2}$ for $^{132}$Sn reasonably match to the experimental levels. Similarly, the four shallowest states, 1$i_{13/2}$, 3$p_{3/2}$, 2$f_{5/2}$, and 3$p_{1/2}$ are also in reasonable agreement with the experimental data for $^{208}$Pb. In lower levels, the calculated values differ significantly with respect to the experimental data. This relative difference in SPEs with respect to experimental data is a common problem non-relativistic and relativistic mean-field calculations \cite{godd13,gorg19} and reference therein. This discrepancy can be connected with the interaction potential and treatment to the spin-orbit interaction in the model. A detailed systematic analysis will be providing light in these directions. In the case of 3$s$- and 2$p$- levels, the calculated ordering of the levels from the NL3$^*$ and DD-ME2 differs from the experimental data for $^{132}$Sn, and $^{208}$Pb, respectively. For example, the 2$d_{3/2}$ level of $^{132}$Sn, and 2$f_{5/2}$ level of $^{208}$Pb, start to be occupied before the 3$s_{1/2}$ and 3$p_{3/2}$ levels, respectively for both the RMF (NL3$^*$) and RHB (DD-ME2) parameter sets, which is in reverse ordering to the experimental data. \\

The right upper and middle panels of Fig. \ref{fig2} display the single-particle energy levels, respectively, for $^{134}$Sn, and $^{210}$Pb nuclei, for both the NL3 (black line), NL3$^*$ (red line), and DD-ME2 (blue line) parameter sets. We did not find any significant difference in the single-particle energy levels for NL3$^*$ and DD-ME2 parameter sets; however, the used pairing descriptions in both are different. This is because the considered nuclei are situated near the $\beta$-stable region of the nuclear landscape. Hence, there is no significant change in using the different prescriptions of pairing correlations. The lower-left and right panel of the Fig. \ref{fig2} shows the SPEs for the rest of the isotopes for the DD-ME2 parameter only, and a similar trend can be expected for NL3 and NL3$^*$ parameter sets. In reference to the $^{132}$Sn ($N = 82$) and $^{208}$Pb ($N = 126$) for the isotopic shift (see Fig. \ref{fig1}), we have studied $^{134}$Sn ($N = 84$), and $^{210}$Pb ($N = 128$) nuclei to determine the relative change (rearrangement) of the single-particle energy levels due to excess neutrons and their effect on the isotopic shift. The lowest neutron single-particle levels for $^{134}$Sn, and $^{210}$Pb are found to be 2$f_{7/2}$ and 2$g_{9/2}$, respectively. Comparing the single-particle levels for $^{132}$Sn and $^{134}$Sn, the addition of two neutrons rearranges the valence levels significantly. As a result, we find partial occupation $(0.236)$ of the nearest higher level 3$p_{1/2}$. Consequently, we get enhancement in the charge radius, which can be noticed in Fig. \ref{fig1}. Further, in the case of $^{210}$Pb, the partial occupation of the 1$i_{11/2}$ level along with the lowest neutron single-particle level 1$g_{9/2}$ modifies the charge radius for all the parameter sets. As a result, we get a sharp isotope shift beyond $N =126$. Our calculated results are in line with the previous predictions with a weaker strength drawn in Ref. \cite{rein17} for Sn-isotopes using the Fayans energy density functional and Ref. \cite{godd13} for the Pb-isotopes using the Skyrme energy density functional. Following these results, it is quite interesting to examine the configuration mixing of orbits near the Fermi surface using the shell model. \\
\\
\noindent
{\bf Shell-model calculations:} The magic numbers are fundamental to nuclear structure physics and were explained using the shell model for the first time in Ref. \cite{mayer55}. To further support our results, we have performed shell model calculations for $^{130}$Sn and $^{134}$Sn, using the Sn100PN and CWG interactions \cite{brown05}, respectively. The Sn100PN interaction assumes $^{100}$Sn as a core. The corresponding neutron single-particle energies are taken to be $-10.6089, -10.2893, -8.7167, -8.6944, -8.8152$ MeV for the 1$g_{7/2}$, 2$d_{5/2}$, 2$d_{3/2}$, 3$s_{1/2}$ and 1$h_{11/2}$ orbits, respectively. The CWG interaction assumes $^{132}$Sn as a core. The active neutron single-particle energies are taken to be $-0.8940, -2.4550, -0.4500, -1.6010, -0.7990, 0.2500$ MeV for the 1$h_{9/2}$, 2$f_{7/2}$, 2$f_{5/2}$, 3$p_{3/2}$, 3$p_{1/2}$ and 1$i_{13/2}$ orbits. The Nushell code of Brown and Rae \cite{brown} has been used to diagonalize the shell model Hamiltonian. The calculated energies are compared to the experimental data in Fig.~\ref{fig3} for the yrast $0^+$ to ${10}^+$, and $0^+$ to ${8}^+$ states, respectively, in $^{130}$Sn and $^{134}$Sn. The calculated energies are in line with the experimental data. All the experimental data have been taken from \cite{nndc}. The chosen interaction is found to work quite well for calculating the energy levels and the electromagnetic quantities \cite{maheshwari2015}. It is to be noted that the single-particle energies are fitted to explain the experimental data around $N = 82$ of the observed proton or neutron energy levels. In Sn100PN interaction, \cite{brown05}, the proton single-particle energies are fitted on the basis of the energy levels observed in $^{133}$Sb \cite{sanchez1998}. Based upon the energy levels observed in $^{131}$Sn \cite{firestone2004}, the neutron single-particle energies are fitted. In CWG interaction, the wave functions for $N >82$ are obtained with the same model space for protons as above and with a model space for neutrons of (1$h_{9/2}$, 2$f_{7/2}$, 2$f_{5/2}$, 3$p_{3/2}$, 3$p_{1/2}$, 1$i_{13/2}$) with respective single-particle energies of $-0.8940, -2.4550, -0.4500, -1.6010, -0.7990,$ and $0.2500$ MeV. Interestingly, the single-particle energies obtained in the RMF are in line with the $spe$ inputs of the shell model.

In the Sn-isotopes, the highest-j, $h$ orbits dominate as $h_{11/2}$ and $h_{9/2}$ just below and above $N = 82$, respectively. In our shell model calculations, the 2$d_{3/2}$, 3$s_{1/2}$ (below $N=82$) and 2$f_{7/2}$, 3$p_{3/2}$ (above $N = 82$) orbits lie in between the $\ell+1/2, h_{11/2}$ and the $\ell-1/2, h_{9/2}$ states. The $d_{3/2}$, $s_{1/2}$ levels are nearly degenerate with the $h_{11/2}$, while a certain energy gap is maintained for the $f_{7/2}$, $p_{3/2}$ orbits with respect to the $h_{9/2}$. The attractive nature of the $\ell+1/2$ orbit can lead to a lower radius in opposite to the repulsive nature of complementary $\ell-1/2$ orbit. This effect becomes prominent for higher $\ell$ values, so for the $h_{11/2}$ and $h_{9/2}$ orbits. A kink is, therefore, expected and has been observed in a recent experimental study \cite{gorg19}.  

The isotopic shift can be defined as the first-order derivative of nuclear charge radii over neutron number $N$, arising via the Coulomb interaction of charged particles (electrons and protons) with the nucleus in the presence of strong nuclear force. Since the shell model Hamiltonian takes care of the proper configuration mixing by using residual interaction, we have also calculated the isotopic shift using occupations of neutrons in the shell model with a description as given in Ref. \cite{caurier01}, as shown in Fig. \ref{fig1}. The isotopic radii variation in Sn isotopes can be written as,
\begin{eqnarray}
\langle \Delta r_{ch}^2 \rangle^A  = \frac{1}{50} n_{h_{11/2}}(A)  b^2 \quad {\bf (below \quad N=82)}, \
\label{smsnshift} \\
\langle \Delta r_{ch}^2 \rangle^A  = \frac{1}{50} n_{f_{7/2}}(A) b^2 \quad {\bf (above \quad N=82)}.
\label{smsnshift1}
\end{eqnarray}
Here, $n_{h_{11/2}}(A)$ and $n_{f_{7/2}}(A)$ are the occupancies of h$_{11/2}$ and $f_{7/2}$ orbits for the Sn isotopes for a given mass number $A$. Also, $b$ is the oscillator size parameter, used as a popular and widely-used prescription of Blomqvist and Molinari \cite{blomqvist68} given by:
\begin{eqnarray}
 b^2  = 0.90 A^{1/3} +0.70 fm^2
\label{oscillator}
\end{eqnarray}
The isotopic shift has been calculated with reference to the doubly magic nuclei, $^{132}$Sn for both $N < 82$ and $N > 82$ regions as given in Eq. (\ref{shift}). The shell model calculated isotopic shift for Sn isotopes from Eqs. (\ref{smsnshift}) and (\ref{smsnshift1}) follow the experimental values quite well. This tells the dominance of h$_{11/2}$ and $f_{7/2}$ orbits in $N \le 82$ and $N \ge 82 $, Sn isotopes, respectively, which is also in line with the origin of seniority isomers in these regions \cite{maheshwari2015, maheshwari2016}. The shell model results explain the kink in charge radii for Sn isotopes and also support the RMF interpretation. To compare the analogy used in Eqs. (\ref{smsnshift}) and (\ref{smsnshift1}), we have also used the occupancies from Relativistic-Hartree-Bogoliubov for DD-ME2 parameter set, and the calculated isotopic shift is shown in Fig. \ref{fig1} as RHB*. The shift again supports a change at closed shell; however, the calculated numbers are a little lesser than the experimental data due to the difference in location of the respective orbit in mean-field calculations.   

On a similar note, we have performed shell model calculations for $^{206}$Pb and $^{210}$Pb by using the Kuo-Herling hole and particle interactions, KHHE and KHPE, respectively \cite{warburton91, mcgrory75, kuo71}. The neutron single-hole energies (below $^{208}$Pb) are taken to be $10.7810, 9.7080, 7.9380, 8.2660, 7.3680, 9.0010$ MeV, respectively, for the 1$h_{9/2}$, 2$f_{7/2}$, 2$f_{5/2}$, 3$p_{3/2}$, 3$p_{1/2}$ and 1$i_{13/2}$ orbits. The neutron single-particle energies (above $^{208}$Pb) are taken to be $-3.1580, -3.9370, -1.4460, -2.3700, -1.4000, -1.9050$, $-2.5140$ MeV, respectively, for the 1$i_{11/2}$, 2$g_{9/2}$, 2$g_{7/2}$, 3$d_{5/2}$, 3$d_{3/2}$, 4$s_{1/2}$, 1$j_{15/2}$ orbits. Fig. \ref{fig4} exhibits the comparison of the experimental and calculated excitation energies versus spin from $0^+$ to ${12}^+$ states in $^{206}$Pb and from $0^+$ to ${10}^+$ states in $^{210}$Pb. The calculated energies reproduce the experimental data reasonably well for both, as shown in Fig. \ref{fig4}. The $g_{9/2}$ orbit becomes active as soon as we cross $^{208}$Pb with increasing neutron number rather than the $i_{11/2}$ orbit. The calculated wave functions from the $0^+$ to $8^+$ states in $^{210}$Pb have major contributions from the $g_{9/2}^2$ configuration. However, the calculated $1^+$ state is found at $1.688$ MeV, arising from the $i_{11/2}^1 g_{9/2}^1$ configuration. The relative energy gap between the $g_{9/2}$ and the $i_{11/2}$ becomes very low which may be understood as the attractive force acting between the $j= \ell-1/2, i_{11/2}$ level and the $j'=\ell'+1/2, g_{9/2}$ with $\Delta \ell=2$ coupling \cite{otsuka05}. Also, the tensor effects will be larger for a higher $\ell$ orbit. The main point of concern is that the $0^+$ ground state in $^{210}$Pb which has significant contribution ($\approx 18 \%$) from the $i_{11/2}^2$ configuration in comparison to the $\approx 66 \%$ contribution from the $g_{9/2}^2$ configuration. This actually highlights the role of radial correlations of the occupied proton $h_{11/2}$ orbit with the unoccupied neutron $i_{11/2}$ orbit due to the same principal quantum number, as discussed by \cite{godd13}. So, the overall structure of this region may be understood by the higher-j, $i$ orbits around $N = 126$. Again, the repulsive nature of the $j=\ell-1/2,i_{11/2}$ orbit, may lead to the higher radius above $N = 126$. The orbits, which lie in between the $i_{13/2}$ and $i_{11/2}$, are the 2$f_{5/2}$, 3$p_{3/2}$, 3$p_{1/2}$ (below $N = 126$), 2$g_{9/2}$ (above $N = 126$). However, the next intruder orbit is $j_{15/2}$ in this valence space. The nature of $i$-orbits around $N=126$, and the nearby location of the next intruder higher-j orbit may result in a large kink in the charge radius.

We have also obtained the isotopic radii variation in Pb isotopes which can be calculated as 
\begin{eqnarray}
\langle \Delta r_{ch}^2 \rangle^A  &=& \frac{1}{82} n_{f_{5/2}\otimes p_{1/2}} (A) b^2 \quad \nonumber \\  && \quad \quad  {\bf (below \quad N=126)},
\label{smpbshift} \\
\langle \Delta r_{ch}^2 \rangle^A  &=& \frac{1}{82} n_{g_{9/2}\otimes i_{11/2}\otimes j_{15/2}} (A) b^2 \quad \nonumber \\  && \quad \quad {\bf (above \quad N=126)}. \label{smpbshift1}
\end{eqnarray}
Here, $n_{f_{5/2} \otimes p_{1/2}}(A)$ and $n_{g_{9/2}\otimes i_{11/2}\otimes j_{15/2}} (A)$ are the occupancies of $f_{5/2} \otimes p_{1/2}$ (below $N=126$) and $g_{9/2}\otimes i_{11/2}\otimes j_{15/2}$ (above $N=126$) orbits for the Pb isotopes for a given mass number $A$. The oscillator size parameter $b$ is defined in Eq. (\ref{oscillator}). The shell model calculated isotope shift reproduces the experimental values below $N=126$, Pb isotopes using Eq. (\ref{smpbshift}) while it could not reproduce the experimental values above $N=126$, Pb isotopes even when the occupancies of many orbits are taken into account as shown in Eq. (\ref{smpbshift1}). This suggests that the shell model calculations need an improvement in terms of used interaction especially in the region above $^{208}$Pb, as shown by Gottardo $et$ $al.$ \cite{gottardo2012}. If we take the total occupancies of $2$, $4$ and $6$ for $^{210}$Pb, $^{212}$Pb and $^{214}$Pb, respectively, beyond $N=126$, the experimental values of isotope shift does not get reproduced. This further hints towards the existence of an extra mechanism resulting in large measured values of charge radii beyond $^{208}$Pb which is taken care by itself in RMF calculations. The isotopic shift using Relativistic-Hartree-Bogoliubov (for DD-ME2 parameter set) occupancies from Eqs. (\ref{smpbshift}) and (\ref{smpbshift1}) (RHB*) also remains far from the experimental data beyond $^{208}$Pb as shown in Fig. \ref{fig1}, while the calculated numbers are quite off from the experimental data below $^{208}$Pb. 

\section{Conclusions}
\label{summary}
\noindent
Concluding, we have presented the isotopic shift in the charge radii of even-even isotopes of Sn and Pb nuclei at $N =  82$ and  $126$,  respectively.  The relativistic mean-field with NL3$^*$ and  Relativistic-Hartree-Bogoliubov approach with DD-ME2 parameter sets are used for the present analysis. The widely used NL3 results are also obtained for standard comparison. The shell model estimation for the isotopic shift in these isotopes is also included. A correlation is established between single-particle levels and charge radii in terms of occupation probabilities to illustrate the shift at $N = 82$ and $126$ for Sn- and Pb- isotopes, respectively. We find a weaker shift for Sn- isotopes as compare to Pb-isotopes, which may be correlated with the softness of Sn-isotopes \cite{biswal15,howa20}. The single-particle energies used in the shell model are consistent with the relativistic mean-field and Relativistic-Hartree-Bogoliubov approach. The shell model calculated isotopic shifts are mostly in line with the RMF results and experimental data. However, the shell model gets limited to explain the isotopic shift for Pb isotopes beyond $N = 126$. The results further support the previous theoretical study of Ref. \cite{godd13} for Pb-isotopes and the recent experimental analysis of Ref. \cite{gorg19} for Sn-isotopes.
 
\section*{Acknowledgments}
\noindent
MB and BVC acknowledge support from FOSTECT Project Code: FOSTECT.2019B.04, FAPESP Project Nos. 2017/05660-0, INCT-FNA Project No. 464898/2014-5, and from the CNPq - Brasil.  PDS acknowledges support from the UK STFC under project number ST/P005314/1. HAK and NY acknowledge support from the Research University Grant (GP0448-2018) under University of Malaya. \\
\\

\end{document}